\documentclass[letterpaper]{article} 
\usepackage[]{aaai2026}  
\usepackage{times}  
\usepackage{helvet}  
\usepackage{courier}  
\usepackage[hyphens]{url}  
\usepackage{graphicx} 
\usepackage{natbib}  
\usepackage{caption} 
\usepackage{amsmath}
\usepackage{algorithm}
\usepackage{algorithmic}
\usepackage{booktabs}
\usepackage{txfonts}
\DeclareSymbolFont{matha}{OML}{txmi}{m}{it}
\DeclareMathSymbol{\varv}{\mathord}{matha}{118}

\usepackage{newfloat}
\usepackage{listings}
\DeclareCaptionStyle{ruled}{labelfont=normalfont,labelsep=colon,strut=off} 
\floatstyle{ruled}
\newfloat{listing}{tb}{lst}{}
\floatname{listing}{Listing}
\usepackage{soul}
\usepackage{xcolor}
\usepackage{graphicx}
\usepackage{subcaption}
\usepackage{xspace}
\usepackage{todonotes}

\usepackage{nicefrac}

\newcommand*\iftodonotes{\if@todonotes@disabled\expandafter\@secondoftwo\else\expandafter\@firstoftwo\fi}  

\title{Triggered: A Statistical Analysis of\\ Environmental Influences on Extremist Groups}
\author{
    Christine de Kock and Eduard Hovy
}
\affiliations{
    University of Melbourne\\

    Parkville, Victoria\\
    3010, Australia\\
    christine.dekock@unimelb.edu.au
}

\usepackage{bibentry}

\begin{document}

\maketitle
\newcommand{\rnews}{r/News\xspace}
\begin{abstract}
Online extremist communities operate within a wider information ecosystem shaped by real-world events, news coverage, and cross-community interaction. We adopt a systems perspective to examine these influences using seven years of data from two ideologically distinct extremist forums (Stormfront and Incels) and a mainstream reference community (\rnews). We ask three questions: how extremist violence impacts community behaviour; whether news coverage of political entities predicts shifts in conversation dynamics; and whether linguistic diffusion occurs between mainstream and extremist spaces and across extremist ideologies. Methodologically, we combine counterfactual synthesis to estimate event-level impacts with vector autoregression and Granger causality analyses to model ongoing relationships among news signals, behavioural outcomes, and cross-community language change. Across analyses, our results indicate that Stormfront and \rnews appear to be more reactive to external stimuli, while Incels demonstrates less cross-community linguistic influence and less responsiveness to news and violent events. These findings underscore that extremist communities are not homogeneous, but differ in how tightly they are coupled to the surrounding information ecosystem.

\end{abstract}

\section{Introduction}
The modern internet is a complex virtual world, consisting of many interconnected social groups that interact and evolve over time. \citet{hill2025catching} remarks that digital spaces function not as isolated environments, but as interconnected emotional ecosystems vulnerable to cross-domain contagion triggered by real-world events. 
In the context of extremist groups, understanding the environmental factors that influence group behaviour is particularly important. 

Prior work has established that real-world extremist violence is often followed by an increase in online hate speech \citep{olteanu2018effect, baele2024diachronic, lupu2023offline}. It can also trigger further acts of violence \citep{kupper2022contagion} and can be used to amplify extremist recruitment attempts \citep{rosenblat2025digital}. 

In addition to being reactive to extremist violence, online communities exist within a broader ecosystem in which minor events and interactions take place daily, all of which may impact upon the group. The influence of such sustained, day-to-day external stimuli is less well-studied. A systematic review by \citet{hassan2018exposure} determined that exposure to radical violent online material is associated with extremist online and offline attitudes at the individual level, but less is known about group-level behaviour, or potential effects of non-violent material.

Prior work has further established that there are complex relationships between different extremist groups: for example, the so-called ``incels'' are a misogynistic extremist group, but their discourse also commonly features racist hate speech \citep{baele2021incel} and their membership has been found to overlap with alt-right groups such as Stormfront \citep{vu2021extremebb}. There is also linguistic evidence of social influence within and across groups. \citet{boholm2025leads} found that extremist hate speech can undergo ``mainstreaming'', whereby general-use platforms follow the adoption of dogwhistles originating in extremist spaces. 
\citet{kupper2022contagion} performed a qualitative investigation of manifestos related to extremist violence in different countries, concluding that there are contagion effects in their language, including intertextual references and shared narratives.    

In this work, our objective is to investigate influence -- defined here as stimuli that evoke measurable responses -- within online extremist groups. \noindent We address the following research questions, each evaluated relative to a mainstream reference community: 
\begin{itemize}
    \item \textbf{RQ1}: Do different online extremist groups respond differently to violent events?
    \item \textbf{RQ2}: How does news coverage of political entities impact conversations in extremist communities?
    \item \textbf{RQ3}: Does language contagion occur between extremist communities with different ideologies?
\end{itemize}

We compare trends over seven years among two related but distinct extremist forums (Incels.is and Stormfront.org, henceforth referred to by domain name only) and one mainstream forum (the \rnews subreddit). Two statistical methods are employed to this end, namely, counterfactual synthesis and Granger causality.

Our results indicate that the Incels forum is comparatively more insulated from external influences. Stormfront is more responsive to both extremist events and to news coverage, and demonstrates more linguistic interaction with both other groups. Out of 36 studied extremist attacks, statistically significant responses are observed for 9 events on both Stormfront and \rnews, with the largest effect being observed for the Oakland Boogaloo shooting (41\% and 129\% increases, respectively, in the number of contributors per day). By comparison, only one event has a significant response on Incels. Increases in news coverage on Donald Trump and Kamala Harris are found to Granger-cause increases in the daily number of posts per user on Stormfront. On \rnews, increases in coverage on George Floyd is found to Granger-cause increases in anger, posts per user, and number of posters. Our results further show evidence of linguistic contagion between \rnews and Stormfront and to a lesser extent between Incels and Stormfront, but not between Incels and \rnews. Together, these results show that external influences can have a real and measurable effect on extremist groups, but that such responses are not homogenous.

\section{Background}
A limited but growing line of research has investigated computational models for extremism. Two works are particularly relevant to ours. 

First, \citet{boholm2025leads} analyse the temporal dynamics of dogwhistles -- concealed hateful expressions -- across two Swedish online groups representing relatively more and less radical communities. Using vector autoregression (VAR) and Granger causality tests, they find that changes in dogwhistle usage in the more radical community precede corresponding changes in the less radical community, whereas the reverse pattern is not observed. This provides evidence of directional linguistic diffusion from more radical to more mainstream spaces. 
Inspired by this approach, we apply VAR and Granger causality to study linguistic diffusion between mainstream and extremist communities, as well as across distinct extremist ideologies (\textbf{RQ3}). We also use the same framework to model the influence of news coverage of different political entities on conversation dynamics (\textbf{RQ2}). For \textbf{RQ3}, our approach further differs from theirs in that (\textit{i)} we analyse shorter time horizons (daily or weekly changes rather than quarterly changes), and (\textit{ii}) we focus on word frequency, whereas they study shifts in word meaning. 

Second, \citet{olteanu2018effect} investigate the effect of violent events on various types of online hate speech. 
They use causal inference methods for this purpose, framing the analysis as measuring the impact of an intervention (the event) on a time series (hate speech volume). A counterfactual time series is constructed based on a structural time series model, which estimates what the series would have looked like if the event had not taken place, following \citet{brodersen2015inferring}. Their findings indicate that extremist violence tends to lead to an increase in messages advocating violence on Reddit and Twitter. 
In this work, we use the same counterfactual synthesis approach to investigate \textbf{RQ1}, replicating the model definition of \citet{olteanu2018effect} (Section \ref{sec:methods}). Our work on RQ1 differs from their study in that (\textit{i}) they study Islamist or Islamophobic attacks, whereas we investigate alt-right and misogynistic attacks, and (\textit{ii}) our scope is larger: we consider 36 attacks over 7 years, whereas they look at 13 attacks over 19 months. Our research questions are also comparatively broader, including multiple signals and directions of influence.

Related analyses by \citet{lupu2023offline} and \citet{baele2021incel} use structural break analysis to identify change points in online activity series. \citet{lupu2023offline} investigate four events across five social platforms, whereas \citet{baele2021incel} study three events across incel-related groups. We opt to use counterfactual synthesis instead, as it more directly supports causal interpretation by constructing an explicit no-event baseline and estimating an effect size with associated uncertainty.

\paragraph{Our contributions}
This work provides a large-scale, longitudinal analysis of how external stimuli impact online extremist communities. We quantify behavioural responses in three groups over seven years, enabling comparisons across communities and with a mainstream reference group. We model multiple environmental stimuli, including news coverage of salient political entities, extra-group linguistic changes, and real-world violence. By estimating effects in three directions (event-to-community, news-to-community and community-to-community), our analyses offer a more holistic account of how extremist groups interact with, and are influenced by, the broader online information ecosystem.

\section{Signal definitions}\label{sec:signals}
\begin{figure*}
    \centering
        \includegraphics[width=\linewidth]{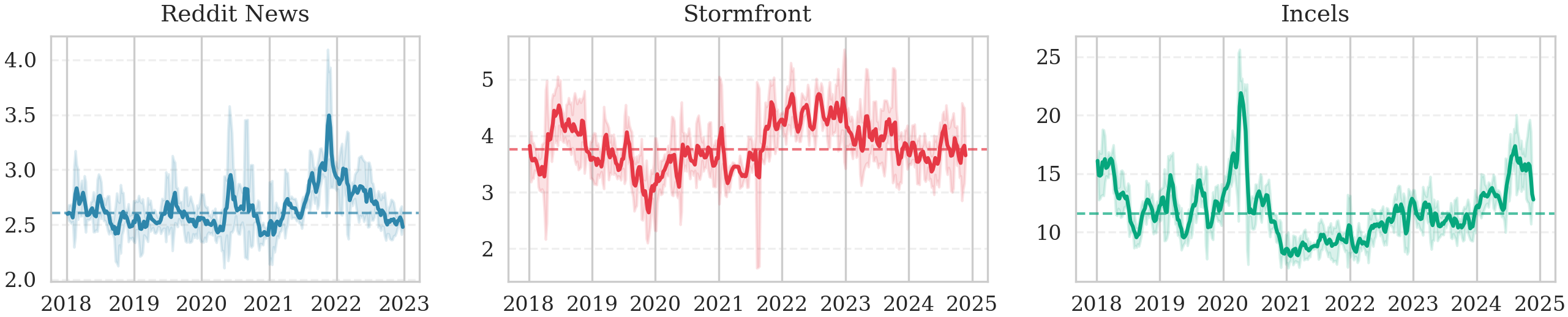}
    \caption{Number of posts per poster over time.}
    \label{fig:posts_per_user}
\end{figure*}

\begin{figure*}
    \centering
    \includegraphics[width=\linewidth]{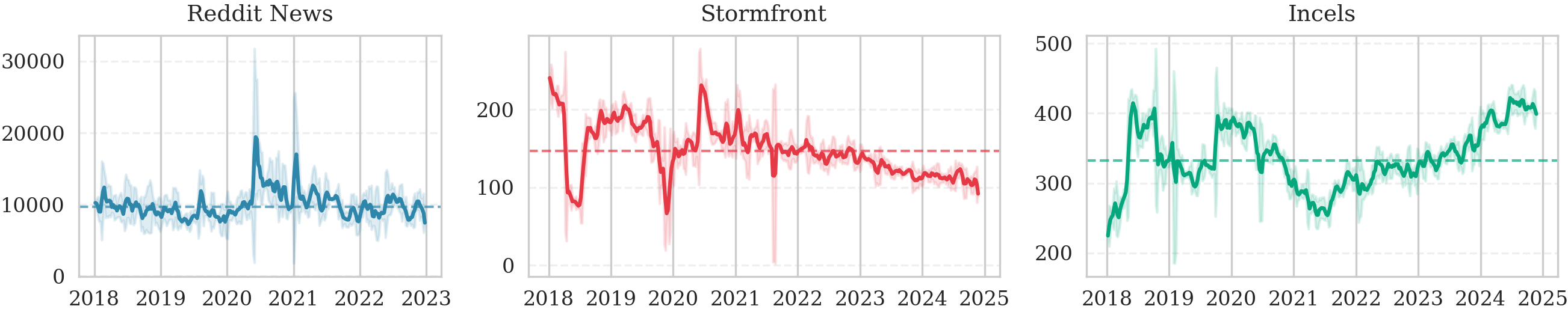}
    \caption{Daily number of posters per platform over time.}
    \label{fig:users}
\end{figure*}

We construct stimulus and response time series to investigate the three research questions, based on two information sources: conversations (Section \ref{sec:sig_convs}) and news (Section \ref{sec:sig_news}). We further construct a set of violent events (Section \ref{sec:sig_events}) and exogenous variables (Section \ref{sec:exogenous}). Section~\ref{sec:methods} details methods used to analyse the relationships between these series.

\subsection{Conversations}\label{sec:sig_convs}
We use the IYKYK dataset \citep{de2025iykyk}, which consists of 19.6 million posts from two online extremist platforms, Stormfront and Incels. Stormfront promotes white supremacy and is one of the oldest online hate groups, whereas Incels is dedicated to misogynistic extremism and was founded in November 2017. We exclude data from before 2018 on either group to remove any initial fluctuations due to platform formation on Incels and ensure a comparable data sample on both platforms. We further use post-2017 data from the \rnews subreddit, collected via the Pushshift API. Due to changes in platform policy, this data has a cutoff of 31 December 2022. The IYKYK and Pushshift datasets are released under a Creative Commons Attribution 4.0 International license. Both datasets consist of only publicly available data, from users who interact on an anonymous basis. The communities in question are primarily English-speaking, with the IYKYK dataset consisting of 95\% English data \citep{de2025iykyk}. This does limit the applicability of our results beyond the English-speaking world; however, the datasets are representative of the two extremist groups of interest.

\newcommand{\numusers}{$\#$posters\xspace}
\newcommand{\postsperuser}{$\#$posts$/$poster\xspace}
\paragraph{Engagement}
\begin{table}[h]
    \centering
    \begin{tabular}{|l|r|r|r|}
    \hline
            & \numusers & \#posts & \postsperuser \\
            \hline
       Incels & 336 & 3 914 & 11.88 \\
       Stormfront & 147 & 563 & 3.82 \\
       \rnews & 9 997 & 26 608 & 2.65 \\
        \hline
    \end{tabular}
    \caption{Mean daily engagement metrics per platform, including number of users who post, number of posts and number of posts per contributor.}
    \label{tab:engagement}
\end{table}

\begin{figure}
    \centering
    \includegraphics[width=0.9\linewidth]{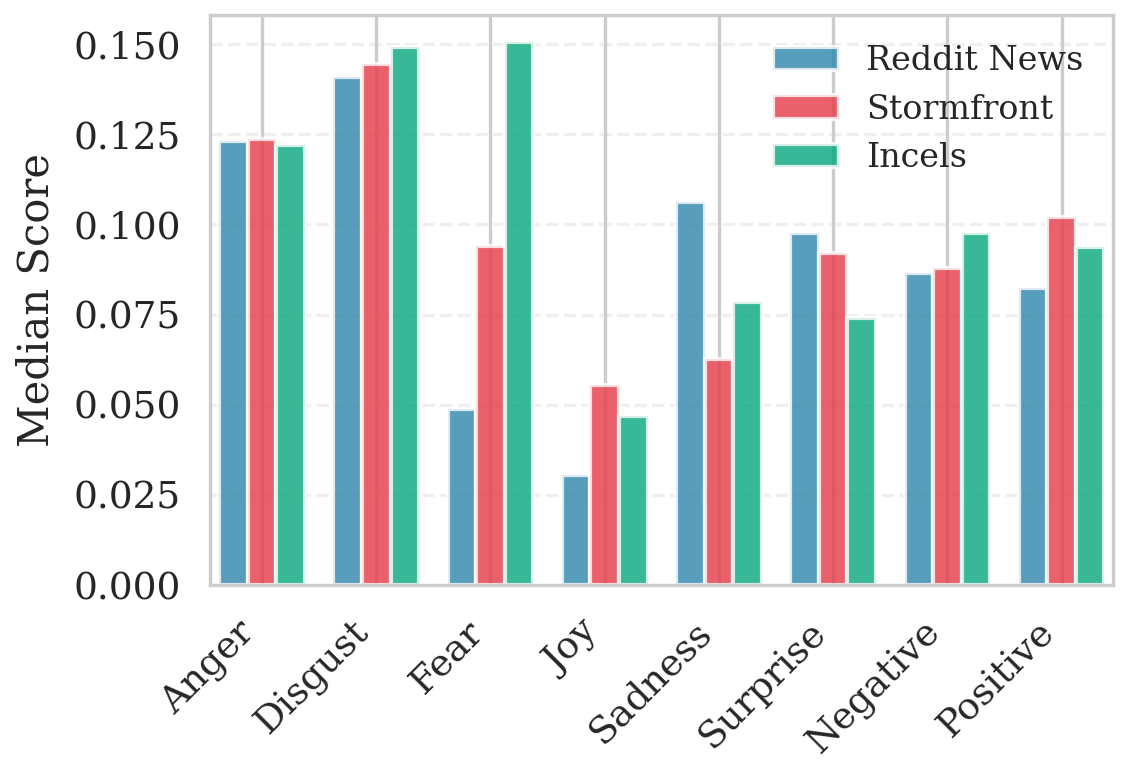}
    \caption{Mean emotion and sentiment levels per platform.}
    \label{fig:emotion_bar}
\end{figure}

\begin{figure*}[t]
\centering
    \includegraphics[width=0.9\textwidth]{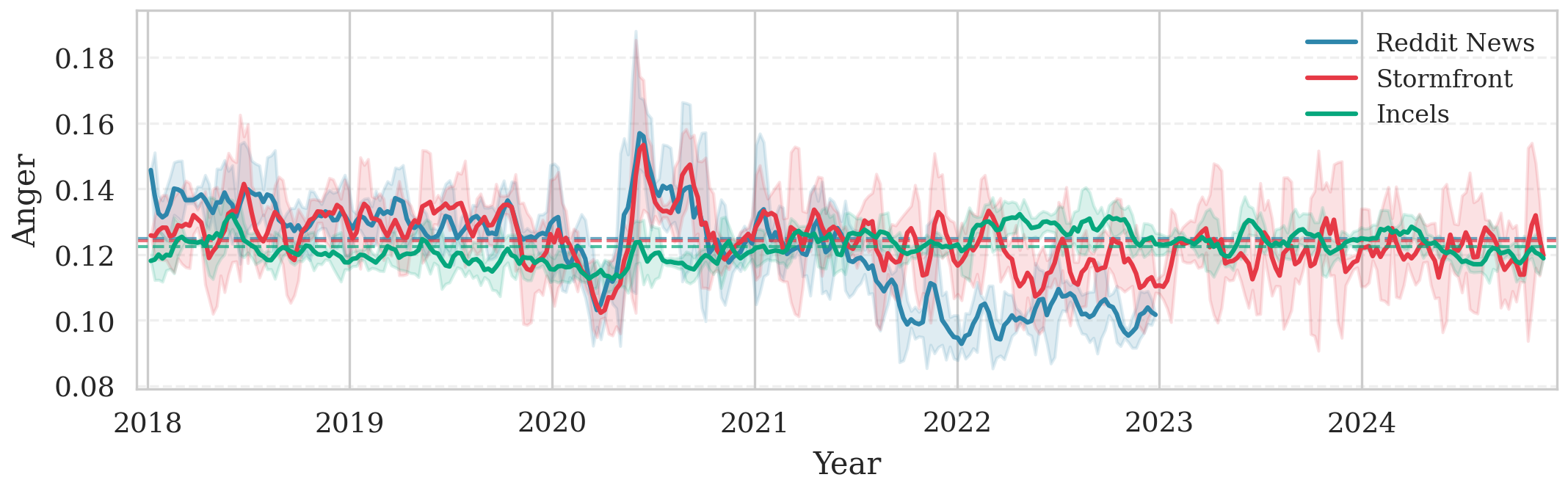}
    \caption{Average daily anger scores across \rnews, Stormfront, and Incels communities (2018-2024).}\label{fig:anger}
\end{figure*}

We extract two daily engagement metrics for each platform: the mean number of users who post (\numusers) and the mean number of posts per user (\postsperuser). 
The \textbf{$\#$users} is considered to be a more robust metric of macro-level engagement than the number of posts, which is susceptible to outliers (e.g. bots). \citet{de2025iykyk} also notes that the number of non-posting users is often up to 100 times more than the active users.

The \textbf{\postsperuser} is intended to capture intensified discussion, which may be limited to a core group. The mean daily values per group are shown in Table \ref{tab:engagement}. Notably, the number of posts in \rnews is substantially larger than on the extremist platforms; however, Incels has a larger number of posts per user than either Stormfront or \rnews, indicating a smaller but highly engaged community. The \numusers and \postsperuser over time are shown in Figure \ref{fig:users} and \ref{fig:posts_per_user}. It is evident that the raw time series are not stationary; in particular, there is an upward trend in the \numusers and \postsperuser on Incels starting early 2021, whereas Stormfront has a downward trend in the daily \numusers over the observation period.

\paragraph{Emotion} To obtain an understanding of affective dimensions of conversation, we also measure the average levels of sentiment and emotions (anger, fear, disgust, sadness, joy, and surprise), using the models of \citet{hartmann2022emotionenglish} and \citet{hutto2014vader}, respectively. The mean intensity values per emotion and sentiment are shown in Figure \ref{fig:emotion_bar}. Of note is that the largest divergence is observed between the scores for fear, with \rnews scoring lower than the extremist groups and Incels scoring the highest. This aligns with prior research in political science, which found that extremist views are often associated with higher levels of fear \citep{van2015fear}. Stormfront scores highest on both positive sentiment and joy, whereas \rnews and Incels have higher levels of sadness, and Incels have more negative sentiment. While the platforms have similar mean levels of anger, the average daily scores (Figure \ref{fig:anger}) show that \rnews and Stormfront underwent a very similar dip and peak in anger in early 2020, corresponding to the COVID-19 pandemic and the Black Lives Matter protests following the death of George Floyd. Prior work by \citet{davies2023witch} found that Stormfront and Incels.co (another incel forum) both experienced increases in post volume during the COVID-19 pandemic. It is also interesting to note that, after the pandemic-related spikes in the \numusers and \postsperuser on Incels, both engagement metrics returned to lower than their pre-pandemic levels by 2021, then rose steadily from there. For \rnews, anger levels seem to have reduced markedly from late 2021, which coincides with spikes in the \postsperuser on both \rnews and Stormfront. 

Our focus on engagement and emotion differs from the approaches of \citet{olteanu2018effect} and \citet{lupu2023offline}, who focus on hate speech in response to extremist violence. This is intentional, as the prior works focused on general-use social media platforms, whereas we specifically research extremist platforms. Therefore, any increase in engagement and emotion is of interest. 

\paragraph{Lexicon}\label{sec:sig_lexicon}
Distinctive language is a common marker of in-group signalling in extremist communities. Following \citet{de-kock-2025-inducing}, we construct community lexicons based on normalized pointwise mutual information (NPMI), which quantifies association between a term \(w\) and community \(c\) beyond independence:
\begin{align}\label{eq:npmi}
    \text{PMI}(w,c) &= \log\frac{P(w\mid c)}{P(w)}, \\
    \text{NPMI}(w,c) &= \frac{\text{PMI}(w,c)}{-\log P(w,c)}.
\end{align}
Probabilities are estimated from token frequencies. A one-vs-all scheme is used: for each community \(c\), \(P(w)\) is computed from the union of the other two communities. To ensure adequate significance, a minimum frequency of 50 usages per community is enforced, and stopwords are removed using the \texttt{nltk} list. 
To ensure adequate temporal support for VAR estimation, we apply a per-pair sparsity filter during analysis, retaining only terms with \(<25\%\) sparsity across the two communities’ weekly series. The top 100 words that survive the sparsity filter are included for each group. Weekly time series are constructed for each of each term, using log-transformed frequencies in the target group. 

\subsection{News}\label{sec:sig_news}
Following \citet{olteanu2018effect}, we use GDELT (Global Data on Events, Location and Tone\footnote{www.gdeltproject.org}), a large open-source dataset documenting news coverage. The GDELT Global Knowledge Graph provides finegrained records of news events, including the relevant entities, locations, publications, and other descriptive codes.

We are interested in how coverage on different political figures influences conversation dynamics. As such, we use the 
daily entity mention counts drawn from the GDELT Global Knowledge Graph. The raw GDELT person entity space contains thousands of names including entertainers, athletes, royalty, locations misclassified as persons, and institutional labels. 
To identify prominent entities, we use the top 50 politically relevant entities by coverage frequency over the observation period, filtering out mistagged person entities (e.g. Asia Pacific, Abu Dhabi) as well as celebrities and fictional entities (e.g. Taylor Swift, Jesus Christ, Harry Potter). The final list of 50 entities is shown in Table \ref{tab:entities}. Time series are constructed based on the log-transformed daily mention count per entity. 

\begin{table}[h]
\centering
\footnotesize
\begin{tabular}{lll}
\hline
Donald Trump & Benjamin Netanyahu & Rahul Gandhi \\
Joe Biden & Mike Pompeo & Tayyip Erdogan \\
Boris Johnson & Ron DeSantis & Kevin McCarthy \\
Vladimir Putin & Hunter Biden & Rudy Giuliani \\
Barack Obama & Imran Khan & Anthony Fauci \\
Kamala Harris & Keir Starmer & Ronald Reagan \\
Narendra Modi & Elizabeth Warren & Harvey Weinstein \\
George Floyd & Bill Clinton & Mark Zuckerberg \\
Hillary Clinton & Gavin Newsom & Jerome Powell \\
Nancy Pelosi & Muhammadu Buhari & Michael Cohen \\
Elon Musk & Andrew Cuomo & Greg Abbott \\
Theresa May & Queen Elizabeth & Pete Buttigieg \\
Bernie Sanders &  Volodymyr Zelenskyy& Jeremy Corbyn \\
Mike Pence & Scott Morrison & Nicola Sturgeon \\
Justin Trudeau & Angela Merkel & John McCain \\
Pope Francis & Antony Blinken & Mitch McConnell \\
Robert Mueller & Brett Kavanaugh & \\
\hline
\end{tabular}
\caption{Entities used in news analysis.}
\label{tab:entities}
\end{table}

\subsection{Violent events}\label{sec:sig_events}
We curate a set of 36 violent events (Table \ref{tab:events} in Appendix \ref{app:events}) that have been credibly linked to alt-right or misogynistic ideologies. First, the \citet{adl_events} documents domestic terror incidents by far-right extremists in the United States from 2017 to 2022, including successful terrorist attacks, failed terrorist attacks and foiled terrorist plots. We filter these to include only attacks that materialised, excluding foiled plots. We further include the events used by \citet{lupu2023offline}, \citet{kupper2022contagion}, \citet{brown2023mis}, \citet{baele2024diachronic} in related studies. The events in \citet{lupu2023offline} include the 2020 US election, the murder of George Floyd, the assassination of General Qasem Soleimani, and a refugee crisis on the borders of Turkey and Greece. Although these events are not directly related to extremist ideologies, they were found to have a significant effect on online hate speech by \citet{lupu2023offline}; as such, we include them in our analysis.

We acknowledge that this set is incomplete and biased towards events in the USA, which stems partly from a lack of credible resources about extremist violence attributed to these ideologies in other parts of the world. Nonetheless, it provides a substantive basis to perform a larger-scale analysis compared to prior work, and a starting point for future work to expand upon.

\subsection{Exogenous variables}\label{sec:exogenous}
Following \citet{olteanu2018effect}, we use exogenous regressors to capture time series variability unrelated to the signals of interest. In particular, we use 
the total daily GDELT news volume (log-transformed, following \citealp{olteanu2018effect}). Since news coverage is influenced by weekly patterns, we further add day-of-week dummy variables.

\section{Methods}\label{sec:methods}
In this work, we apply two methods, both originating from econometrics, to model relationships between the signals discussed in Section \ref{sec:signals}. Vector autoregression, in combination with Granger causality testing, is used to analyse continuously varying external stimuli in the form of news coverage and language influences. Counterfactual synthesis is used to estimate the effect of extremist attacks on conversation dynamics, by modelling these events as interventions.

\subsection{VAR modelling and Granger causality}\label{sec:bg_var}
Vector autoregression (VAR, \citealp{sims1980macroeconomics}) is a method for modelling temporal dependencies between time series, used to analyse how the present value of a series depends on its own previous (\textit{lagged}) values and the lagged values of other series Let $y_t$ denote a vector consisting of two endogenous variables ($k=2$), representing the values of the stimulus and response series as observed at time $t$. We use the VAR-X($p$) formulation, with exogenous controls $x_t$, as defined as \citep{lutkepohl2005stable}:
\begin{equation}
y_t = \nu + A_1y_{t-1} + \cdots + A_py_{t-p} + Bx_t + u_t,
\end{equation}
where $A_q$ is a $k\times k$ matrix of autoregressive coefficients at lag $q$, $p$ is the total number of lags considered, $\nu$ is a vector of $k$ intercept terms, $B$ is a $k\times m$ matrix of coefficients for the $m$ exogenous variables in $x_t$, and $u_t$ is an error term. 

The exogenous variables include day-of-week dummies and log-transformed news volume at time $t$, as described in Section \ref{sec:exogenous}, and are only used for the news coverage VAR analysis. For the language analysis, we follow \citet{boholm2025leads} in using only the endogenous series. 

The error term $u_t$ is assumed to be a vector white-noise process, and model adequacy is assessed via diagnostics applied to the estimated residuals. The lag order $p$ is selected empirically. Following \citet{boholm2025leads}, we use a combination of the Akaike information criterion (AIC; \citealp{akaike2003new}), Bayesian information criterion (BIC; \citealp{schwarz1978estimating}), and Hannan–Quinn information criterion (HQIC; \citealp{hannan1979determination}), computed for the VAR system likelihood, to determine the optimal lag for each model. When the criteria disagree, the majority choice is used; in the absence of agreement, the AIC-optimal lag is selected (per \citealp{boholm2025leads}). 

\subsubsection{Granger causality}
A variable $X$ is said to Granger-cause variable $Y$ if past values of $X$ provide statistically significant information for forecasting $Y$ beyond that contained in $Y$'s own history \citep{box2014time}. Within the VAR framework, Granger causality is tested by comparing two VAR models: a restricted model regressing $Y$ on its own lags, and an unrestricted model including lagged values of both $Y$ and $X$. If the unrestricted model yields a statistically significant improvement in fit based on an F-statistic, past values of $X$ are said to Granger-cause $Y$. Of note is that Granger causality measures temporal precedence rather than mechanistic causation. Observed relationships may reflect the influence of unobserved common factors affecting both series \citep{box2014time}. As such, results must be interpreted cautiously. Nonetheless, these methods have been used similarly in prior work \citep{boholm2025leads} and are utilized here (\textit{i}) in combination with other statistical analyses, and (\textit{ii}) in a comparative rather than absolute manner, to identify relative differences across communities.

\subsubsection{Implementation} We use the Python \texttt{statsmodels} implementation of the VAR-X model. We follow the preprocessing steps outlined in \citet{boholm2025leads}, including imputation, differencing, and stationarity testing. Last-observation-carried-forward (LOCF) imputation is applied prior to differencing to preserve calendar alignment and ensure comparability of lagged observations. Stationarity is required for valid inference in VAR models \citep{lutkepohl2005stable}. In our case, the endogenous variables are often not stationary, as observed from the trends in the \numusers in Section \ref{sec:signals}; as such, differencing is applied to remove stochastic trends. For count-based responses (unique users, post volume, and word frequency), the natural logarithm is applied and the resulting series are first-differenced. Sentiment and emotion measures, as well as \postsperuser, are directly first-differenced. Results must therefore be interpreted in terms of changes in the underlying endogenous signals rather than their absolute levels; for instance, modelling whether a \textit{change} in news coverage on a particular politician precedes a \textit{change} in engagement. 

For each entity--forum--metric combination, the differenced stimulus and response series are subjected to standard unit root tests using the Augmented Dickey-Fuller (ADF, \citealp{dickey1979distribution}) and Kwiatkowski-Phillips-Schmidt-Shin (KPSS; \citealp{kwiatkowski1992testing}) procedures. Series are retained only if the ADF test rejected the unit-root null at the 5\% level while the KPSS test did not reject the stationarity null at the same level, consistent with \citet{boholm2025leads}. 
False discovery rate (FDR) correction is applied using the \citet{benjamini1995controlling} procedure at $q=0.05$ at the level of forum and response pair -- for instance, when testing negative sentiment response on Stormfront, we apply the correction over the set of negative sentiment results for the 50 news entities. Both directions are evaluated, meaning that $n=100$.

\subsection{Counterfactual synthesis}\label{sec:bg_cf}
Our second technique is used to infer the effect of a once-off event, or intervention, on an outcome metric. Prior work has proposed that, in the absence of randomised control trials, a synthesised counterfactual time series may be used to represent 
the expected behaviour had the intervention not occurred. The method of \citet{brodersen2015inferring} models the counterfactual using structural time series models, and was similarly used by \citet{olteanu2018effect} to quantify the effect of 13 Islamophobic and Islamist attacks on online hate speech. The counterfactual is constructed using (\textit{i}) the pre-intervention behaviour of the target series and (\textit{ii}) other time series that predicted the target series before the intervention. The causal effect is estimated from the difference between the observed and counterfactual values.

Concretely, the structural time series model decomposes the observed series into additive components \citep{harvey2005}:
\begin{equation}
y_t = \mu_t + \gamma_t + \beta' x_t + \epsilon_t,
\end{equation}
where $\mu_t$ is a local linear trend component, $\gamma_t$ represents seasonal effects, $\epsilon_t$ is an observation error, and $\beta' x_t$ captures regression effects of predictor variables. The predictor vector $x_t$ at time $t$ includes lagged values from the target series at specific prior time points, as well as contemporaneous exogenous controls at time $t$ (further details below). A Kalman filter is used to estimate how each component (trend, seasonality, predictors) contributes to the observed series in the pre-intervention period. The learned relationships are then extrapolated in the post-intervention period to generate counterfactual forecasts $\hat{y}_t$. The causal effect is quantified as the \textbf{relative effect}, i.e. the cumulative difference between observed values $y_t$ and counterfactual predictions $\hat{y}_t$ over $d$ days post-intervention, normalised by the total observed activity: 
\begin{equation}
\text{Relative Effect} = \frac{\sum_{t=1}^{d} (y_t - \hat{y}_t)}{\sum_{t=1}^{d} y_t}.
\end{equation}

 To ensure that results are robust to false positives, we use a conservative threshold and only consider events to have had a significant impact if the 99\% confidence interval of the accumulated effect does not include 0 (further discussed in Section \ref{sec:res_events}). 

\paragraph{Implementation} In this analysis, our aim is to measure impact on conversation dynamics in online spaces in the week following an extremist attack. We replicate the implementation of \citet{olteanu2018effect}, using maximum likelihood estimation to fit the model with the \texttt{statsmodels} \texttt{UnobservedComponents} Python class. Four data sources are used to construct the counterfactual:
\begin{itemize}\itemsep0em
    \item the observed series in the 11 weeks prior to the event, 
    \item the observed series exactly one year earlier, for 12 weeks (i.e. the 11 weeks before and 1 week after the event),
    \item the observed series 23 weeks prior to the event, for 12 weeks, and 
    \item external information, in the form of day-of-the-week dummy variables and total news volume.
\end{itemize}

The model is fitted on the 11-week pre-intervention period, modelling how $y_t$ at each time point relates to lagged values from one year prior and 23 weeks prior, along with the exogenous variables at $t$. To enable counterfactual forecasting for the week following the event, the historical series extend one week beyond the pre-intervention window, providing a 12-week span at each lag.

Since multiple events are tested, we treat each platform-signal combination as a separate research question and perform multiple testing correction over the 36 events per grouping.

\begin{table*}[h!]
\centering
\footnotesize
\begin{tabular}{llll}
\hline
\textbf{Event} & \textbf{Date} & \textbf{Response} & \textbf{Effect} \% \\
\hline
\multicolumn{4}{c}{\textbf{Incels}} \\
\hline
Nashville school shooting & 2023-03-27 & posts/user & 25.9*** \\
\hline
\multicolumn{4}{c}{\textbf{\rnews}} \\
\hline
El Paso Walmart attack & 2019-08-03 & users & 38.7*** \\
\hline
Assassination of General Qasem Soleimani & 2020-01-03 & users & 46.1*** \\
\hline
Santa Cruz County Boogaloo shootout & 2020-05-06 & posts/user & 17.7*** \\
\hline
Murder of George Floyd & 2020-05-25 & posts/user & 15.9*** \\
 &  & users & 124.7*** \\
 \hline
Oakland Boogaloo shooting & 2020-05-29 & posts/user & 23.5*** \\
 &  & users & 129.3*** \\
 \hline
Kyle Rittenhouse shooting & 2020-08-25 & posts/user & 25.8*** \\
\hline
US Elections & 2020-11-03 & users & 74.2** \\
\hline
US Capitol riot & 2021-01-06 & users & 121.4*** \\
\hline
Uvalde school shooting & 2022-05-24 & users & 63.4** \\
\hline
\multicolumn{4}{c}{\textbf{Stormfront}} \\
\hline
New Zealand mosque attack & 2019-03-15 & users & 19.5*** \\
\hline
Murder of George Floyd & 2020-05-25 & posts/user & 30.8*** \\
 &  & users & 20.9*** \\
 \hline
Oakland Boogaloo shooting & 2020-05-29 & posts/user & 27.6*** \\
 &  & users & 41.0*** \\
 \hline
Kyle Rittenhouse shooting & 2020-08-25 & users & 17.1*** \\
\hline
US Elections & 2020-11-03 & users & 23.2** \\
\hline
US Capitol riot & 2021-01-06 & posts/user & 39.6*** \\
 &  & users & 26.1*** \\
 \hline
Buffalo supermarket mass shooting & 2022-05-14 & users & 13.4** \\
 \hline
Moscow Crocus City Hall attack & 2024-03-22 & users & 11.5** \\
\hline
Donald Trump assassination attempt & 2024-07-13 & posts/user & 28.1*** \\
 &  & users & 22.2*** \\
\hline
\end{tabular}
\caption{Statistically significant results for the effect of violent events on engagement in online groups. $^{**}p<.01$, $^{***}p<.001$.}
\label{tab:event_effects}
\end{table*}

\subsection{Experimental setup}\label{sec:experiments}
In this section, we detail how the methods in Section \ref{sec:bg_var} and \ref{sec:bg_cf} are used to model the signals in Section \ref{sec:signals}. Our experiments are structured in three parts, corresponding to our three research questions.

\paragraph{RQ1: Acute impact of violent events}
To estimate the short-term effect of extremist violence and other salient offline events (Section \ref{sec:sig_events}), we model each event as a once-off intervention and apply counterfactual synthesis. For each platform, we evaluate effects on log-transformed number of unique daily contributors (\numusers) and intensity of discussion among contributors (\postsperuser). Each event is analysed over a fixed seven-day post-intervention window to capture immediate community reaction while limiting contamination from longer-term drift and unrelated events.

\paragraph{RQ2: Continual impact of news coverage}
We model relationships between news attention and community behaviour using bivariate VAR models with Granger causality tests. The stimulus series is daily GDELT entity visibility for the 50 political entities (Section \ref{sec:sig_news}). For each forum and each community outcome metric (engagement and affective measures from Section~\ref{sec:sig_convs}), we estimate a separate VAR-X model for each entity--metric pairing, including day-of-week indicators and total news volume as exogenous controls. To focus inference on short-run responsiveness while limiting over-parameterisation, we restrict candidate lag horizons to the preceding two weeks (i.e., at most 14 daily lags), and select the final lag order per model using information criteria (AIC/BIC/HQIC) as described in Section~\ref{sec:methods}. 
We evaluate both directions (\textit{news$\rightarrow$community} and \textit{community$\rightarrow$news}) to distinguish responsiveness from feedback loops. 

\paragraph{RQ3: Continual impact of language diffusion}
To test whether linguistic diffusion occurs across ideologically distinct communities, we again use VAR/Granger analysis but at weekly granularity to reduce sparsity and align with slower-moving lexical adoption dynamics. 
The log-transformed weekly lexicon counts are modelled in bivariate VAR systems. Here we restrict candidate lag horizons to the preceding six weeks (i.e., at most 6 weekly lags) to capture plausible diffusion delays without diluting power on sparse signals; the final lag order is selected per model using information criteria (as per Section \ref{sec:bg_var}). As with news coverage, we evaluate both directions for each term and control for multiple testing within each community-pair family.

\section{Impact of violent events}\label{sec:res_events}
\begin{figure*}[htbp]
\centering
\begin{subfigure}[b]{0.49\textwidth}
    \centering
    \includegraphics[width=\textwidth]{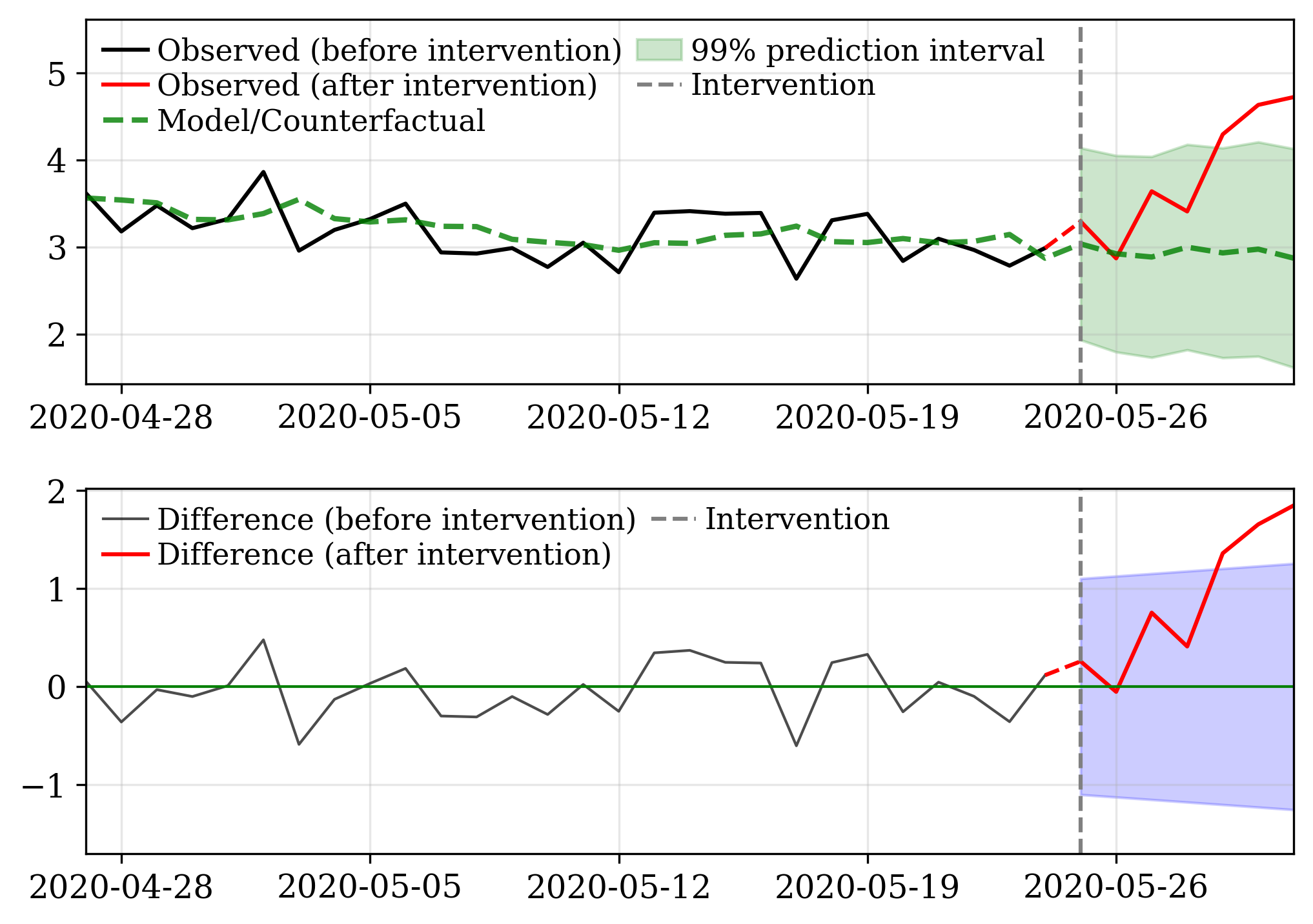}
    \caption{Murder of George Floyd (25-05-2020).}
    \label{fig:sub1}
\end{subfigure}
\hfill
\begin{subfigure}[b]{0.49\textwidth}
    \centering
    \includegraphics[width=\textwidth]{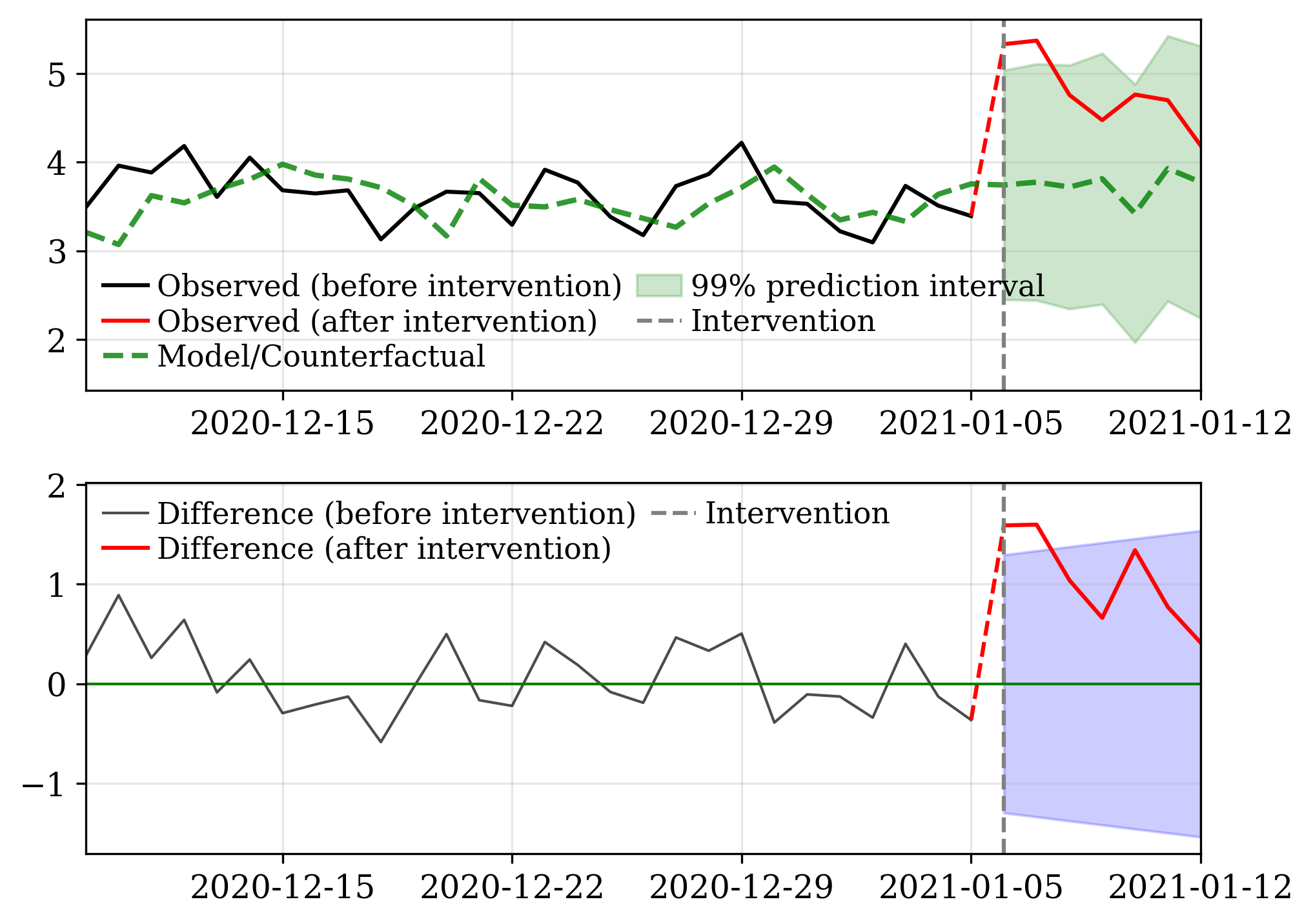}
    \caption{January 6 US Capitol Attack (06-01-2021).}
    \label{fig:sub2}
\end{subfigure}
\caption{Impact of real-world events on \postsperuser on Stormfront. Top panels show the observed and modelled values. Bottom panels show the difference between these values. Shaded areas indicate 99\% prediction intervals.}
\label{fig:events_combined}
\end{figure*}

Results for the counterfactual synthesis analysis are shown in Table~\ref{tab:event_effects}, reporting only event--signal pairs exhibiting statistically significant post-event deviations.

For Stormfront, 9 out of 36 events have a significant impact (13 event--signal pairs). Three events affect both \numusers and \postsperuser, indicating both increased participation and higher posting intensity among active users. The largest estimated effects are observed for the Oakland Boogaloo shooting (29~May~2020; 41.0\% for \numusers and 27.6\% for \postsperuser). Notably, this event occurred four days after the murder of George Floyd, meaning that their post-event windows overlap, and effect sizes may be inflated for one or both events. This highlights a caveat, which is that the method operates solely on dates and is not able to deconfound multiple events in close temporal proximity. 

Figure~\ref{fig:events_combined} visualises two Stormfront examples for \postsperuser. In both cases, the counterfactual closely matches pre-event dynamics, with residual differences centred near zero. Post-event, the George Floyd series shows a sustained upward deviation relative to the counterfactual (consistent with the abovementioned overlapping effects) whereas the January~6 Capitol attack yields a sharper, more immediate increase within the 7-day window.

For \rnews, 9 out of 26 events have significant effects (11 event--signal pairs), with the largest effect again observed for the Oakland Boogaloo shooting (129.3\% for \numusers), plausibly interacting with the contemporaneous George Floyd effect. Stormfront and \rnews share five significant events; however, \rnews results are only available through 2022, meaning that more recent Stormfront-relevant events (e.g., the 2024 Trump assassination attempt) cannot be evaluated for \rnews. 

For Incels, only one event out of 36 is significant (the Nashville school shooting; 27~March~2023). This is somewhat surprising, given that there are several other attacks in our events set that have been directly linked to this group. A potential explanation is that there can be delays in attacks being attributed to a specific ideology, particularly when the target is not immediately obvious. While alt-right attacks often target places of worship, the incel attacks in our events set took place in a public street (Toronto, Plymouth), a shopping center (Glendale), and a massage parlour (Atlanta). A delayed community response would not be captured by our method, which focuses only on the 7 days following an attack. Alternatively, effects may manifest primarily as topical shifts rather than changes in posting volume.

Two of the four events from \citet{lupu2023offline} have a significant impact on both \rnews and Stormfront based on our intervention analysis: the 2020 US elections, and the murder of George Floyd. The assassination of the Iranian General Qasem Soleimani further had a significant effect on \rnews, but not for the two extremist groups. This highlights that current affairs beyond extremist violence can influence online (extremist) groups. In the Section \ref{sec:granger_news}, this idea is explored further, by investigating their responses to news coverage.

\paragraph{Validation}\label{sec:events_val}
Given the sensitivity of the domain,  
we validated that our method does not produce excessive false positives. We conducted a placebo analysis by placing pseudo-events three weeks before each actual event and estimating their effect. 
Using identical parameters to the main analysis, we analysed all dates in our events set for which the required lag data is available (35 pseudo-events for Stormfront and Incels, and 25 for \rnews). 
Under the null hypothesis, these dates should show no significant effects -- however, we note that genuine significant events could feasibly occur on these dates as well, potentially yielding conservative estimates of false positive rates (FPR). In fact, there is a notable concentration of four events from 19-29 May 2020, which are expected to yield such ``false'' false positives. 

Table \ref{tab:fpr_analysis} shows the number of false positives obtained when using a 95\% vs 99\% confidence interval to determine statistical significance (with multiple testing correction applied). We observe a maximum false positive rate of 5.7\% (or two events) at 99\%\, with the events cluster in May 2020 indeed yielding both significant pseudo-events. By comparison, an FPR of 20\% (7 events) is observed for 95\%.  Given our preference to prioritise precision, our results discussion above 
is based on the more conservative threshold.

\section{Impact of entity news coverage}\label{sec:granger_news}
\begin{table}[htbp]
\centering
\footnotesize
\setlength{\tabcolsep}{3pt}
\begin{tabular}{lrl}
\hline
\textbf{Entity--Response} & \textbf{F} & \textbf{Significant lags} \\
\hline
\multicolumn{3}{c}{\textbf{\rnews} (\textit{News coverage $\rightarrow$ community})} \\
\hline
George Floyd--anger & $2.45^{**}$ & $1^{+}$--$7^{+}$ (\nicefrac{7}{14}) \\
George Floyd--\postsperuser & $4.55^{***}$ & $1^{+}-6^{+},13^{-},14^{-}$ (\nicefrac{8}{14}) \\
George Floyd--\numusers & $4.95^{***}$ & $1^{+}-6^{+}$ (\nicefrac{6}{6}) \\
Pete Buttigieg--anger & $2.65^{**}$ & $4^{+}$ (\nicefrac{1}{14}) \\
Pete Buttigieg--sentiment\textcolor{red}{$^{\dagger\dagger}$} & $4.40^{***}$ & $3^{-},5^{-}$ (\nicefrac{2}{9}) \\
\hline
\multicolumn{3}{c}{\textbf{\rnews} (\textit{Community $\rightarrow$ news coverage})}\\
\hline
George Floyd--\numusers & $3.52^{**}$ & $1^{+}-6^{+}$ (\nicefrac{6}{6}) \\
\hline
\multicolumn{3}{c}{\textbf{Stormfront} (\textit{News coverage $\rightarrow$ community})} \\
\hline
Donald Trump--\postsperuser& $4.70^{***}$ & $1^{+}-5^{+}$ (\nicefrac{5}{6}) \\
Kamala Harris--\postsperuser & $2.68^{**}$ & $1^{+},6^{+}-8^{+}$ (\nicefrac{4}{12}) \\
\hline
\end{tabular}
\caption{Statistically significant Granger relationships for news impact on community behaviour ($^{*}$ = $p<.05$, $^{**}= p<.01$, $^{***} = p<.001$.). Daggers indicate Ljung-Box test failure and lag superscripts indicate coefficient direction. $\nicefrac{x}{p}$ indicates the number of individually significant lag coefficients out of $p$ lags in the selected VAR.}
\label{tab:granger_news_effects}
\end{table}

Table \ref{tab:granger_news_effects} contains the significant results ($p<0.05$, with multiple testing correction) for the Granger causality analysis of news entity coverage. Seven relationships achieve statistical significance. Five of these are observed on \rnews, where increases in coverage of George Floyd Granger-cause increases in anger ($F=2.45, p=.009$), \postsperuser ($F=4.55, p<.001$), and \numusers ($F=4.95, p<.001$). Bidirectional effects were observed for \numusers ($F=3.52, p=.002$ for reverse direction), with community activity also predicting subsequent news coverage. This pattern suggests a feedback loop in which news coverage drives online engagement, which in turn amplifies subsequent news attention. All other significant relationships are in one direction only, with news coverage leading community responses. 

Two significant relationships are observed on Stormfront. Increases in news coverage of Donald Trump ($F=4.70, p<.001$) and Kamala Harris ($F=2.68, p=.013$) are associated with increases in \postsperuser. No significant results are observed for Incels.

\paragraph{Validation}
In addition to pre-testing the transformed series as described in Section \ref{sec:bg_var}, three diagnostic criteria are used to assess significant Granger relationships: Ljung-Box tests for residual autocorrelation, and ADF and KPSS tests for residual stationarity. All models achieved residual stationarity on both ADF and KPSS tests. 
All but one of the significant relationships pass all Ljung-Box tests, indicating well-specified models with no remaining autocorrelation. 
\textit{Pete~Buttigieg$\leftrightarrow$\rnews negative sentiment} failed diagnostics with significant autocorrelation in both stimulus and response residuals (marked with $\dagger\dagger$). Such bidirectional diagnostic failures indicate remaining unmodelled time dependence due to misspecification or shared shocks affecting both series, meaning that the lead-lag ordering cannot be interpreted reliably.

Examining the lag coefficient structure, the George Floyd relationships show sustained immediate effects with most significant lags concentrated in the first 7 days and in the positive direction: all 7 tested lags for anger (lags $1^{+}$ -- $7^{+}$), all 6 tested lags for \numusers (lags $1^{+}$ -- $6^{+}$), and 6 of the first 7 lags for \postsperuser (lags $1^{+}$ -- $6^{+}$), with additional delayed negative effects emerging at lags $13^{-}$, $14^{-}$. This indicates robust immediate lagged dynamics. \textit{Donald~Trump$\rightarrow$Stormfront} \textit{\postsperuser} similarly exhibits sustained immediate effects with 5 consecutive significant lags (lags $1^{+}$ -- $5^{+}$ of 6 tested). \textit{Kamala~Harris$\rightarrow$Stormfront \postsperuser} shows a more dispersed pattern with 4 of 12 lags significant (lag $1^{+}$ and lags $6^{+}$ -- $8^{+}$). In contrast, both Pete Buttigieg relationships show sparse coefficient patterns (1 of 14 lags for anger, 2 of 9 lags for sentiment) with sign reversals in the sentiment relationship, consistent with the diagnostic failure noted above.

\section{Cross-community language diffusion}\label{sec:res_language}
\begin{table}[t]
\centering
\footnotesize
\begin{tabular}{lll}
\hline
\textbf{N-gram} & \textbf{F} & \textbf{Significant lags} \\
\hline
\multicolumn{3}{c}{\textit{\rnews $\rightarrow$ Incels}} \\
\hline
protected\textcolor{red}{$^{\dagger}$} & 10.10$^{**}$ & -- (\textcolor{red}{\nicefrac{0}{3}}) \\
\hline
\multicolumn{3}{c}{\textit{Stormfront $\rightarrow$ Incels}} \\
\hline
jewish\textcolor{red}{$^{\dagger}$} & 9.26$^{*}$ & -- (\textcolor{red}{\nicefrac{0}{2}}) \\
jews\_in & 4.29$^{*}$ & $3^{+},4^{+}$ (\nicefrac{2}{6}) \\
\hline
\multicolumn{3}{c}{\textit{\rnews $\rightarrow$ Stormfront}} \\
\hline
incident\textcolor{red}{$^{\dagger}$} & 10.10$^{*}$ & -- (\textcolor{red}{\nicefrac{0}{2}}) \\
enforcement\textcolor{red}{$^{\dagger\dagger}$} & 8.60$^{*}$ & -- (\textcolor{red}{\nicefrac{0}{2}}) \\
law\_enforcement\textcolor{red}{$^{\dagger\dagger}$} & 8.33$^{*}$ & $1^{+}-2^{+}$ (\nicefrac{2}{2}) \\
protests & 7.50$^{**}$ & $1^{+}$ (\nicefrac{1}{4}) \\
armed & 7.04$^{*}$ & $1^{+}-2^{+}$ (\nicefrac{2}{4}) \\
moderator & 6.12$^{*}$ & $1^{+}$ (\nicefrac{1}{4}) \\
brutality & 5.83$^{**}$ & $1^{+}-6^{+}$ (\nicefrac{6}{6}) \\
cdc & 5.74$^{*}$ & $1^{+},3^{+}$ (\nicefrac{2}{4}) \\
the\_police & 5.38$^{*}$ & $1^{+}-4^{+}$ (\nicefrac{4}{4}) \\
their\_job\textcolor{red}{$^{\dagger}$} & 5.18$^{*}$ & $1^{+}-2^{+},4^{+}$ (\nicefrac{3}{4}) \\
correction & 4.98$^{*}$ & $5^{-}$ (\nicefrac{1}{5}) \\
\hline
\multicolumn{3}{c}{\textit{Stormfront $\rightarrow$ \rnews}} \\
\hline
flu\textcolor{red}{$^{\dagger}$} & 7.75$^{*}$ & -- (\textcolor{red}{\nicefrac{0}{3}}) \\
italy\textcolor{red}{$^{\dagger}$} & 7.20$^{*}$ & $2^{+}-3^{+}$ (\nicefrac{2}{3}) \\
planes\textcolor{red}{$^{\dagger}$} & 5.74$^{*}$ & $1^{+}-2^{+}$ (\nicefrac{2}{4}) \\
\hline
\end{tabular}
\caption{Statistically significant Granger causality results for vocabulary diffusion across online platforms. ($^{*}$ = $p<.05$, $^{**}= p<.01$, $^{***} = p<.001$.). Daggers indicate Ljung-Box test failure and lag superscripts indicate coefficient direction. $\nicefrac{x}{p}$ indicates the number of individually significant lag coefficients out of $p$ lags in the selected VAR.}
\label{tab:granger_vocab_diffusion}
\end{table}

Table~\ref{tab:granger_vocab_diffusion} reports statistically significant Granger causality relationships ($p<.05$) between weekly first-differenced word frequencies across online communities. The largest set of significant results occurs between \rnews and Stormfront (14 terms). Most relationships in this group flow from \rnews to Stormfront, particularly for terms related to public events, institutions, and policing (e.g., \texttt{protests}, \texttt{armed}, \texttt{cdc}, \texttt{the\_police}). A smaller number of terms show the reverse pattern, including \texttt{flu}, \texttt{italy}, and \texttt{planes}. A more limited but thematically concentrated set of relationships are found between Incels and Stormfront, with significant effects indicated for \texttt{jewish} and \texttt{jews\_in}. 

The significant terms are often thematically related to news events, particularly for \textit{\rnews$\leftrightarrow$Stormfront}, and less related to ideological or in-group terminology, with the exception of \texttt{jews} and \texttt{jews\_in}. This may be due to the sparsity of group-specific language in out-group communities. Prior work has found that Incels exhibits significantly more highly specialised in-group language compared to Stormfront \citep{de2025iykyk} -- for instance, inventing words like \texttt{tallcel} to refer to an incel who is tall -- which is highly unlikely to be used in the other two groups.
However, this also means that we cannot exclude the possibility that the groups are responding to the same external stimuli, with Stormfront generally being slightly slower to respond, rather than there being a directional inter-group influence. However, either explanation would support a tighter linguistic coupling between \rnews and Stormfront, compared to either group with Incels.

In contrast to \citet{boholm2025leads}, the dominant direction observed is from a mainstream to an extreme group (\textit{\rnews$\rightarrow$Stormfront}). However, our study differs from \citet{boholm2025leads} in that (\textit{i}) they specifically focus on in-group language, (\textit{ii}) they look at longer timeframes (one year versus six weeks), and (\textit{iii}) their metrics relate to changes in word meaning as opposed to word frequency. While their results provide evidence of mainstreaming of extremist dogwhistles, our results indicate that there is a lead-lag relationship from the mainstream to the extreme group in their discussion of topical events.

\paragraph{Validation}
All reported results pass residual stationarity tests (ADF and KPSS). Ljung-Box diagnostics, however, indicate residual autocorrelation in 9 of 17 models: seven models show autocorrelation in one residual series (\texttt{protected}, \texttt{jewish}, \texttt{flu}, \texttt{italy}, \texttt{planes}, \texttt{their\_job}, \texttt{incident}) and two show autocorrelation in both series (\texttt{enforcement}, \texttt{law\_enforcement}). Table~\ref{tab:granger_vocab_diffusion} reports the number of individually significant lags alongside the joint F-statistics, indicating substantial heterogeneity in lag structure. A number of terms show sustained effects, with a majority of individual lags reaching significance: \texttt{brutality} (6/6 lags), \texttt{the\_police} (4/4 lags), \texttt{their\_job} (3/4 lags), and \texttt{law\_enforcement} (2/2 lags). By contrast, five terms pass the joint Granger causality test despite having no individually significant lag coefficients:
\texttt{protected} (0/3), \texttt{jewish} (0/2), \texttt{flu} (0/3), \texttt{incident} (0/2), and \texttt{enforcement} (0/2).
This suggests the joint Granger result reflects weak, distributed effects across multiple lags rather than a dominant effect at any single lag. Combined with residual autocorrelation failures (all five cases violate the Ljung--Box test), these results should be treated cautiously, as they may reflect misspecified dynamics or shared shocks rather than stable lead--lag structure.

\section{Discussion}\label{sec:discussion}

A clear pattern emerges across the three analyses in this work: \rnews and Stormfront illustrate more responsiveness to news coverage and events, substantial overlap in their significant events (5 out of 9 events), and more inter-platform linguistic interaction. The Incels forum, by contrast, appears to be less responsive to both news and events, and shows limited cross-community linguistic interaction. These results suggest that this group may be less receptive to environmental influences, indicating a more insulated and internally driven conversational dynamic. A possible explanation for this result may be that their ideology is focused on their own inability to form romantic relationships with women, which implies a more introspective focus. Such insularity is also consistent with the finding that significantly more in-group language is used on the Incels platform \citep{de2025iykyk}. By contrast, Stormfront self-identifies as a white nationalist community, which naturally entails engagement with societal issues such as political events, racial incidents, and immigration policy -- topics that dominate mainstream news coverage and thus create substantial overlap with the content discussed on \rnews. Stormfront and \rnews represent the smallest and the largest groups by volume, based on Table \ref{tab:engagement}; as such, these results are not likely to be confounded by group size. Furthermore, we observe these patterns in metrics that are independent of group size, such as the average negative sentiment, \postsperuser and anger levels. 

Two entities emerge as significant across the event and news analyses: George Floyd and Donald Trump. News coverage on Donald Trump is associated with increases in \postsperuser on Stormfront, and the January 6 US Capitol riot and the 2020 US elections had a significant impact on Stormfront and \rnews. The attempted assassination of Donald Trump also had a significant impact on Stormfront and was not evaluated for \rnews due to the data cut-off date. An increase in news coverage on George Floyd is associated with increases in anger, \postsperuser and \numusers, and the date of his murder is a significant event for both \rnews and Stormfront. This complements the findings of \citet{lupu2023offline}, who observed spikes in hate speech following this event across several online platforms.

Though the objective of this work is to characterise the factors that influence extremist groups, our results indicate that \rnews is more responsive to news coverage than either of the extremist communities and has the same number of extremist attacks with significant responses as Stormfront. This is perhaps not surprising, given that the focus of \rnews is solely on news, whereas the extremist groups are primarily focused on their respective ideologies. Based on this insight, we can conclude that extremist groups do not uniformly respond to extremist violence, and significant responses to such violence are not confined to extremist groups.

\section*{Ethical Statement}
\label{sec:ethics}
This work analyses aggregate relationships between external stimuli (news, language, and offline events) and online community behaviours. The goal is to improve understanding of how communities respond to the wider information environment, which we believe to be an important question. However, we recognise that this is a contentious topic and that it is important to proceed with appropriate caution. We recognise that there is a risk that findings may be misinterpreted or overstated. In particular, Granger causality assesses temporal predictability, not mechanistic causation; nonetheless, results may be read as causal claims or used to support simplistic narratives about particular communities. Such misreadings can contribute to stigmatisation, sensational media framing, or policy overreach (e.g., broad censorship or surveillance) that affects benign users. To mitigate these risks, we (\textit{i}) restrict reporting to community-level, time-aggregated analyses, (\textit{ii}) apply conservative inference procedures (diagnostic checks and false-discovery control), (\textit{iii}) frame results as descriptive evidence about system-level dynamics rather than guidance for identifying or acting on individual users, and (\textit{iv}) repeatedly highlight caveats and uncertainty. By presenting three complementary analyses (event-level impacts, news--behaviour dynamics, and cross-community language diffusion), we triangulate evidence across methods, reducing reliance on any single model assumption or isolated association.

\section{Conclusion}
In this work, we use Granger causality analysis and synthetic control methods to investigate the influence of news coverage, extremist attacks, and cross-community linguistic diffusion on Stormfront, Incels, and \rnews. Our results indicate substantial differences in responsiveness across the three communities. Both \rnews and Stormfront exhibit strong responsiveness to political violence and news coverage, with substantial overlap in events and extensive linguistic interaction. Incels, by contrast, shows minimal responsiveness to events and limited cross-platform linguistic diffusion. These patterns align with ideological foci: Stormfront's white nationalist ideology engages with political events and news cycles that dominate mainstream coverage, while Incels' inward focus on personal romantic grievances shows little connection to external political developments. Notably, the mainstream news-oriented platform (\rnews) proves more responsive to extremist violence than either extremist community, demonstrating that responsiveness to such events is not exclusive to extremist spaces.%

\bibliography{aaai2026}

\newcommand{\answerYes}[1]{\textcolor{blue}{#1}} 
\newcommand{\answerNo}[1]{\textcolor{teal}{#1}} 
\newcommand{\answerNA}[1]{\textcolor{gray}{#1}} 
\newcommand{\answerTODO}[1]{\textcolor{red}{#1}}

\section{Checklist}

\begin{enumerate}
\item For most authors...
\begin{enumerate}
    \item  Would answering this research question advance science without violating social contracts, such as violating privacy norms, perpetuating unfair profiling, exacerbating the socio-economic divide, or implying disrespect to societies or cultures?
    \answerYes{Yes, this research advances our understanding of online communities violating social contracts. Care is taken to be careful about interpretations, to highlight caveats, and to avoid false positive results.}
  \item Do your main claims in the abstract and introduction accurately reflect the paper's contributions and scope?
    \answerYes{Yes, all claims are supported by statistical tests and results in Sections \ref{sec:res_events}, 
    \ref{sec:granger_news} and \ref{sec:res_language}.}
   \item Do you clarify how the proposed methodological approach is appropriate for the claims made? 
    \answerYes{Yes, Section  and \ref{sec:experiments}.}
   \item Do you clarify what are possible artifacts in the data used, given population-specific distributions?
    \answerYes{Yes. We note a US-focused bias (Section \ref{sec:events_val}) and the fact that discussions are primarily in English (Section \ref{sec:sig_convs}).}
  \item Did you describe the limitations of your work?
    \answerYes{Yes, caveats and limitations are discussed throughout the paper.}
  \item Did you discuss any potential negative societal impacts of your work?
    \answerYes{Yes, Section \ref{sec:ethics}.}
      \item Did you discuss any potential misuse of your work?
    \answerYes{Yes, Section \ref{sec:ethics}.}
    \item Did you describe steps taken to prevent or mitigate potential negative outcomes of the research, such as data and model documentation, data anonymization, responsible release, access control, and the reproducibility of findings?
    \answerYes{Yes, we use conservative reporting thresholds and provide clear instructions on cautious interpretations of the results.}
  \item Have you read the ethics review guidelines and ensured that your paper conforms to them?
    \answerYes{Yes, we have read and understood these guidelines.}
\end{enumerate}

\item Additionally, if your study involves hypotheses testing...
\begin{enumerate}
  \item Did you clearly state the assumptions underlying all theoretical results?
    \answerYes{Yes. Our assumptions are primarily related to temporal frames within which changes are observed. The implications of these choices are discussed in detail.}
  \item Have you provided justifications for all theoretical results?
    \answerYes{Yes, results have been tied to known differences in ideologies in Section \ref{sec:discussion}.}
  \item Did you discuss competing hypotheses or theories that might challenge or complement your theoretical results?
    \answerYes{Yes. We note that results may be due to temporal precedence, feedback cycles or co-occurence.}
  \item Have you considered alternative mechanisms or explanations that might account for the same outcomes observed in your study?
    \answerYes{Yes, as above.}
  \item Did you address potential biases or limitations in your theoretical framework?
    \answerYes{Yes. We note the effect of community size, US/English bias, and inability to distinguish between temporal co-occurrence and causal impact.}
  \item Have you related your theoretical results to the existing literature in social science?
    \answerYes{Yes, we results have been related to prior work on extremism.}
  \item Did you discuss the implications of your theoretical results for policy, practice, or further research in the social science domain?
    \answerYes{Yes, we note the important implication of not treating these groups as homogenous.}
\end{enumerate}

\item Additionally, if you are including theoretical proofs...
\begin{enumerate}
  \item Did you state the full set of assumptions of all theoretical results?
    \answerNA{N/A}
	\item Did you include complete proofs of all theoretical results?
    \answerNA{N/A}
\end{enumerate}

\item Additionally, if you ran machine learning experiments...
\begin{enumerate}
  \item Did you include the code, data, and instructions needed to reproduce the main experimental results (either in the supplemental material or as a URL)?
    \answerNA{N/A}
  \item Did you specify all the training details (e.g., data splits, hyperparameters, how they were chosen)?
    \answerNA{N/A}
     \item Did you report error bars (e.g., with respect to the random seed after running experiments multiple times)?
    \answerNA{N/A}
	\item Did you include the total amount of compute and the type of resources used (e.g., type of GPUs, internal cluster, or cloud provider)?
    \answerNA{N/A}
     \item Do you justify how the proposed evaluation is sufficient and appropriate to the claims made? 
    \answerNA{N/A}
     \item Do you discuss what is ``the cost`` of misclassification and fault (in)tolerance?
    \answerNA{N/A}
  
\end{enumerate}

\item Additionally, if you are using existing assets (e.g., code, data, models) or curating/releasing new assets, \textbf{without compromising anonymity}...
\begin{enumerate}
  \item If your work uses existing assets, did you cite the creators?
    \answerYes{Yes, Section  \ref{sec:sig_convs} and \ref{sec:sig_news}.}
  \item Did you mention the license of the assets?
    \answerYes{Yes, Section \ref{sec:sig_convs} and \ref{sec:sig_news}.}
  \item Did you include any new assets in the supplemental material or as a URL?
    \answerNA{We do not create any new assets.}
  \item Did you discuss whether and how consent was obtained from people whose data you're using/curating?
    \answerYes{Yes, Section \ref{sec:sig_convs} mentions that the data is publicly available.}
  \item Did you discuss whether the data you are using/curating contains personally identifiable information or offensive content?
    \answerYes{Yes, Section \ref{sec:sig_convs} mentions that the data is anonymous and from extremist groups.}
\item If you are curating or releasing new datasets, did you discuss how you intend to make your datasets FAIR?
\answerNA{No new datasets are released.}
\item If you are curating or releasing new datasets, did you create a Datasheet for the Dataset? 
\answerNA{No new datasets are released.}
\end{enumerate}

\item Additionally, if you used crowdsourcing or conducted research with human subjects, \textbf{without compromising anonymity}...
\begin{enumerate}
  \item Did you include the full text of instructions given to participants and screenshots?
    \answerNA{N/A}
  \item Did you describe any potential participant risks, with mentions of Institutional Review Board (IRB) approvals?
    \answerNA{N/A}
  \item Did you include the estimated hourly wage paid to participants and the total amount spent on participant compensation?
    \answerNA{N/A}
   \item Did you discuss how data is stored, shared, and deidentified?
   \answerNA{N/A}
\end{enumerate}

\end{enumerate}

\appendix
\section{List of extremist events}\label{app:events}
The complete list of extremist events is shown in Table \ref{tab:events}.

\begin{table}[htbp]
\centering
\small
\caption{Events analyzed in the study}\label{tab:events}
\begin{tabular}{ll}
\hline
Date & Event \\
\hline
15-03-2019 & New Zealand mosque attack \\
27-04-2019 & Synagogue attack in Poway, California \\
17-06-2019 & Dallas courthouse incel attack \\
19-11-2019 & Gainesville church attack \\
03-08-2019 & El Paso Walmart attack \\
10-08-2019 & Norway Mosque attack \\
09-10-2019 & Germany synagogue attack \\
03-01-2020 & Assassination of Gen. Soleimani \\
24-02-2020 & Toronto incel attack \\
28-02-2020 & Turkey / Greece refugee crisis \\
31-03-2020 & Los Angeles train attack \\
20-05-2020 & Glendale, Arizona incel attack \\
29-05-2020 & Oakland Boogaloo shooting \\
25-05-2020 & George Floyd murder \\
06-05-2020 & Santa Cruz County boogaloo shootout \\
19-05-2020 & Courthouse attack and murders \\
25-08-2020 & Kyle Rittenhouse shooting \\
03-11-2020 & US Elections \\
06-01-2021 & US Capitol riot \\
16-03-2021 & Atlanta, USA incel attack \\
26-06-2021 & White supremacist attack in Wintrop, Mass. \\
19-08-2021 & Racist school attack in Sweden \\
12-08-2021 & Plymouth incel shooting \\
14-05-2022 & Buffalo supermarket mass shooting \\
24-05-2022 & Uvalde school shooting \\
19-11-2022 & Colorado Springs LGBTQ club attack \\
21-01-2023 & Shooting at the Star Ballroom Dance Studio \\
27-03-2023 & Nashville school shooting \\
03-05-2023 & Belgrade school shooting \\
06-05-2023 & Allen Texas shooting \\
25-10-2023 & Lewiston, Maine shooting \\
21-12-2023 & Prague university attack \\
22-03-2024 & Moscow Crocus City Hall attack \\
13-04-2024 & Bondi mall stabbing \\
15-04-2024 & Wakeley (Australia) mass stabbing \\
13-07-2024 & Attempted assassination of Donald Trump \\
\hline
\end{tabular}
\end{table}

\section{False positive analysis}
The results for the false positive analysis are shown in Table \ref{tab:fpr_analysis}.
\begin{table}[htbp]
\centering
\footnotesize
\begin{tabular}{lcc}
\hline
Metric & 95\% FPR & 99\% FPR \\
\hline
\multicolumn{3}{c}{\textbf{Stormfront (n=35)}} \\
\hline
\postsperuser  & 11.4\% & 2.9\% \\
\numusers  & 20.0\% & 5.7\% \\
\hline
\multicolumn{3}{c}{\textbf{Incels (n=35)}} \\
\hline
\postsperuser & 0.0\% & 0.0\% \\
\numusers  & 8.6\% & 5.7\% \\
\hline
\multicolumn{3}{c}{\textbf{\rnews (n=25)}} \\
\hline
\postsperuser  & 8.0\% & 4.0\% \\
\numusers  & 0.0\% & 0.0\% \\
\hline
\end{tabular}
\caption{False positive analysis results}\label{tab:fpr_analysis}
\end{table}

\end{document}